\documentclass[useAMS,usenatbib]{mn2e}
\usepackage{graphicx} 
\usepackage{times}

\def\kms{km ${\rm s}^{-1}$}

\def\ch2{$\chi^2$}

\def\kms {\hbox{${\rm km\ s}^{-1}$}}

\def\ccm {$\hbox{{\rm cm}}^{-3}$}    
\def\scm  {$\hbox{{\rm cm}}^{-2}$}    

\def\MOLH {\hbox{${\rm H}_2$}}  

\def \AL {$\alpha $}     
\def \HI {H{\sc \,i}}

\def\lapp{\ifmmode\stackrel{<}{_{\sim}}\else$\stackrel{<}{_{\sim}}$\fi}
\def\gapp{\ifmmode\stackrel{>}{_{\sim}}\else$\stackrel{>}{_{\sim}}$\fi}
\def\bsp_small{\vspace{0.5cm}\small\noindent This paper has been typeset
from a \TeX/\LaTeX\ file prepared by the author.\normalsize}

\title[\HI\ 21-cm absorption searches in DLAs]{New searches for \HI\ 21-cm in damped Lyman-{\boldmath $\alpha$}
absorption systems} 
\author[S. J. Curran et
al.]{S. J. Curran$^{1}$\thanks{E-mail: sjc@phys.unsw.edu.au},
P. Tzanavaris$^{2,3,4}$, J. K. Darling$^{5}$, M. T. Whiting$^{6}$,
J. K. Webb$^{1}$, \newauthor
C. Bignell$^{7}$, 
R. Athreya$^{8}$ and  M. T. Murphy$^{9}$\\ 
$^{1}$School of Physics, University of New South Wales, Sydney NSW 2052, Australia\\
$^{2}$Institute of Astronomy and Astrophysics, National Observatory of Athens, I. Metaxa \& V. Paulou
152 36 Penteli, Greece\\
$^{3}$Department of Physics and Astronomy, Johns Hopkins University, Baltimore, MD 21218, USA\\
$^{4}$Laboratory for X-Ray Astrophysics, NASA Goddard Space Flight Center, Code 662, Greenbelt, MD 20771, USA\\
$^{5}$Department of Astrophysical and Planetary Sciences, University of Colorado at Boulder, Boulder, Colorado 80309, USA\\
$^{6}$CSIRO Australia Telescope National Facility, PO Box 76, Epping NSW 1710, Australia\\
$^{7}$National Radio Astronomy Observatory, P.O. Box 2, Rt. 28/92
Green Bank, WV 24944-0002, USA\\
$^{8}$National Centre for Radio Astrophysics, Pune 411 007, Maharashtra, India\\ 
$^{9}$Centre for Astrophysics and Supercomputing, Swinburne University
of Technology, PO Box 218, Hawthorn, VIC 3122, Australia\\
}
\begin{document}

\date{Accepted ---. Received ---; in original form ---}

\pagerange{\pageref{firstpage}--\pageref{lastpage}} \pubyear{2009}

\maketitle

\label{firstpage}

\begin{abstract}
We present the results of three separate searches for \HI\ 21-cm
absorption in a total of twelve damped Lyman-$\alpha$ absorption
systems (DLAs) and sub-DLAs over the redshift range $z_{\rm
    abs}=0.86-3.37$. We find no absorption in the five systems for
which we obtain reasonable sensitivities and add the results to those
of other recent surveys in order to investigate factors which could
have an effect on the detection rate: We provide evidence that the mix
of spin temperature/covering factor ratios seen at low redshift may
also exist at high redshift, with a correlation between the 21-cm line
strength and the total neutral hydrogen column density, indicating a
roughly constant spin temperature/covering factor ratio for all of the
DLAs searched. Also, by considering the geometry of a flat expanding
Universe together with the projected sizes of the background radio
emission regions, we find, for the detections, that the 21-cm line
strength is correlated with the size of the absorber. For the
non-detections it is apparent that larger absorbers (covering factors)
are required in order to exhibit 21-cm absorption, particularly if
these DLAs do not arise in spiral galaxies. We also suggest that the
recent $z_{\rm abs} = 2.3$ detection towards TXS 0311+430 arises in a
spiral galaxy, but on the basis of a large absorption cross-section
and high metallicity, rather than a low spin temperature
\citep{ykep07}.
\end{abstract}


\section{Introduction}\label{intro}

Damped Lyman-$\alpha$ absorption systems (DLAs) are believed to be the
precursors of modern day galaxies, containing at least 80\% of the
neutral gas mass density of the Universe \citep{phw05}. As the name
suggests, DLAs are identified through their heavily damped absorption
features, due to the large columns of neutral hydrogen ($N_{\rm
  HI}\geq2\times10^{20}$ \scm) through which the background quasar is
viewed.  The 21-cm spin-flip transition is of interest since it traces
the cool component of the gas, with the comparison of the 21-cm
absorption strength to the total neutral hydrogen column giving the
gas spin temperature ($T_{\rm spin}$) for a fully absorbed emission
region ($f=1$) [see Equ.~\ref{enew}, Sect. \ref{or}]. Searches for
21-cm absorption in DLAs exhibit a $\approx50\%$ detection rate, 
  the detections occuring predominately at low redshift ($z_{\rm
  abs}\lapp1$), suggesting an increase in the $T_{\rm spin}/f$
ratio with redshift, which could be due to higher spin temperatures
and/or lower covering factors.

In order to shed light on which is the predominant factor, we have
undertaken searches with the Green Bank (GBT) and Giant Metrewave
Radio (GMRT) Telescopes. In particular, we aim to:
\begin{enumerate}
  \item Test the hypothesis of \citet{cmp+03} that 21-cm absorption
    should be readily detectable at high redshift, despite the large
    $T_{\rm spin}/f$ ratios, by searching for 21-cm absorption in DLAs
    which lie towards compact radio sources.
\item Test the hypothesis of \citet{ctp+07} that the spin temperature
  does not continue to increase with redshift, by searching in previously
  unsearched high redshift ($z_{\rm abs}\gapp3.2$) DLAs.
    \item Test the line strength--metallicity correlation ($T_{\rm
      spin}/f$\,--\,[M/H] anti-correlation) of \citet{ctp+07}, which
      suggests that several, currently undetected, DLAs should be
      readily detectable in 21-cm absorption.
\end{enumerate}
We have observed a dozen DLAs and sub-DLAs and in this
paper we present and discuss the results in the context of the above
issues.

\section{Sample Selection and Observations}\label{obs}

\subsection{GBT 2006 observations}
\label{GBT06}
\subsubsection{Motivation}

Prior to the recent high redshift detections
(\citealt{kse+06,kcl06,ykep07}, discussed below), all of the DLAs
detected in 21-cm absorption were at redshifts of $z_{\rm abs}\lapp2$,
where there is also an approximately equal number of
non-detections. \citet{kc02}, interpreted this as the low redshift DLAs
having a mix of spin temperatures, whereas the high redshift absorbers
have exclusively high spin temperatures. However, all of
the $z_{\rm abs}\gapp1$ absorbers have large DLA-to-QSO angular
diameter distance ratios ($DA_{\rm DLA}/DA_{\rm QSO}\approx1$),
whereas the $z_{\rm abs}\lapp1$ absorbers have a mix of ratios, due to
the geometry effects of a flat expanding Universe \citep{cw06}: Those
with low ratios are almost always detected in 21-cm absorption,
whereas those with high ratios tend not to be detected, which is how
the ``spin temperature'' distribution of \citet{kc02} is
segregated. This strongly suggests, on the grounds of geometry alone
and not allowing for various absorber and emitter sizes, that the
covering factor, rather than the spin temperature, is the crucial
criterion in determining the detectability of 21-cm absorption, where
an absorber at high redshift will always cover the background quasar
much less effectively than an equivalent absorber--quasar system at
low redshift. Furthermore, \citet{cmp+03} had previously suggested
that the non-detections at high redshift could be due to low covering
factors, as opposed to elevated spin temperatures, and so 21-cm
absorption should be readily detectable in high redshift DLAs towards
very compact radio sources.

With the Green Bank Telescope we therefore embarked upon a survey for
\HI\ 21-cm absorption in the known radio-illuminated DLAs yet to be
searched, prioritised by a small radio source size. For this we used
the Sloan Digital Sky Survey Damped Ly{$\alpha$} Survey Data Release 1
\citep{ph04} combined with the previously unsearched systems
\citep{cwbc01}, giving 57 DLAs occulting radio-loud ($S\gapp0.1$ Jy)
quasars. We then short-listed the DLAs and sub-DLAs in which the 21-cm
transition was redshifted into the Prime Focus 342 and 450 MHz bands
and which were located towards the four most compact radio sources --
B3~0758+475 (SDSS J080137.67+472528.2), PKS~1402+044, PKS B1418--064
and PKS~2136+141 (Table \ref{size}, where we also give the radio
source sizes for the other objects searched).
\begin{table}
\begin{center}
\caption{The radio source sizes at $\nu_{\theta}$ of the QSOs
  illuminating the DLAs searched (Table \ref{res}) used for selecting the 
  GBT 2006 targets. These were
  obtained from the highest resolution radio images available, closest
  in frequency to the redshifted 21-cm values ($\nu_{\theta}$), although
  0454+039, 0528--250 and 1418--064 have since been imaged at closer
  frequencies \citep{klm+09}. For
  the remainder of the DLA sample refer to \citet{cmp+03,ctp+07}.
 \label{size}}
\begin{tabular}{@{}l c c  c @{}} 
\hline
QSO & $\theta_{\rm QSO}$ [arc-sec] & $\nu_{\theta}$ [GHz] & Reference \\
\hline
0454+039   &  $<0.4$  &  2.3 & \citet{bgp+02} \\
0528--250 &    0.010  &  5.0 & \citet{ffp+00}  \\
0758+475 &   $0.91$ &  1.4 & FIRST\\ 
J0816+4823 &   $0.35$ &  1.4 & FIRST\\ 
0957+561A  &   11.36  &  5.0  &    \citet{hsmr97}          \\
1402+044   &  1.35          &  1.4 &  FIRST \\
1418--064   &  1.27          &  1.4 &   FIRST\\
2136+141   &  0.002         &  5.0 &  \citet{gkf99}\\
\hline
\end{tabular}
\end{center}
{FIRST: The Very
Large Array's {\it Faint Images of the Radio Sky at Twenty
Centimetres}.}
\end{table}

\subsubsection{Observations and data reduction}

The observations were performed with the GBT PF1 342 and
450~MHz receivers in position switching mode (level-3 sampling, one
spectral window and dual polarisation). The backend was either the
GBT spectrometer using 32\,768 lags over a 12.5~MHz band, giving a
channel spacing of 381.47~Hz, or when RFI appeared to be particularly
bad, the spectral processor, using 1024 lags over a 10~MHz band, giving
a channel spacing of 9.766 kHz.


0758+475 was observed
on the 1st of December 2006 for a total of 7 hours. At a central
frequency of 336 MHz, the channel spacing of the GBT spectrometer gave
a resolution of 0.34 \kms.  The system temperature was typically
$T_{\rm sys}\approx120$ K and RFI was not
\begin{figure}
 \centering
\includegraphics[angle=270,scale=0.63]{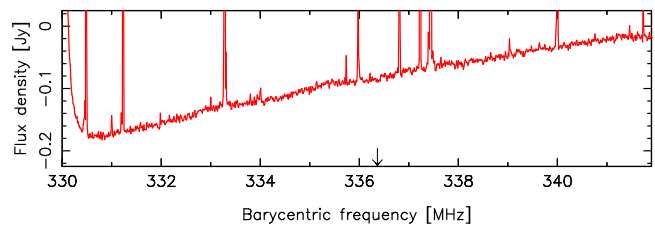}
\caption{GBT 336 MHz band spectrum towards 0758+475 shown at a
  velocity resolution of 10 \kms.  The arrow in this and the following
  spectra (Figs. \ref{1402_385} to \ref{0957+561}) indicates the expected frequency of
  the absorption.}
\label{0758+475}
    \end{figure}
too severe, allowing 6.0 hours of data to be retained. Although most
spectral baselines were flat with the odd intermittent spike, the
measured fluxes were found to vary considerably between scans (from
$\approx-1$ to $\approx1$ Jy), not permitting us to determine the flux
density of this source (Fig.~\ref{0758+475}).

1402+044 was observed at a central frequency of 385 MHz on the 30th of
November and 1st of December 2006, for a total of 1.7 hours.  As per
previous  WSRT observations towards this object (described in \citealt{ctp+07}),
the relatively wide band of the GBT spectrometer could cover the
$z=2.688$ (385.14 MHz) and $z=2.713$ (382.55 MHz) candidate sub-DLAs
of \citet{twl+89}, as well as the $z=2.708$ DLA (383.07 MHz).  At 385
MHz, the GBT spectrometer gives a spectral resolution of 0.30~\kms.
\begin{figure}
 \centering
\includegraphics[angle=270,scale=0.63]{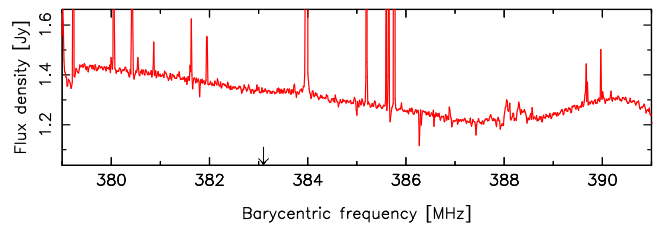}
\caption{GBT 385 MHz band spectrum towards 1402+044 shown at a velocity resolution of 10 \kms.}
\label{1402_385}
    \end{figure}
Typically, the system temperatures were $T_{\rm sys}\approx110$ K and
the removal of RFI dominated scans left a total observing time of 1.4
hours (Fig. \ref{1402_385}).

The $z=2.485$ absorber was observed using the spectral processor,
tuned to a central frequency of 408 MHz, on the 6th and 8th of January
\begin{figure}
 \centering
\includegraphics[angle=270,scale=0.63]{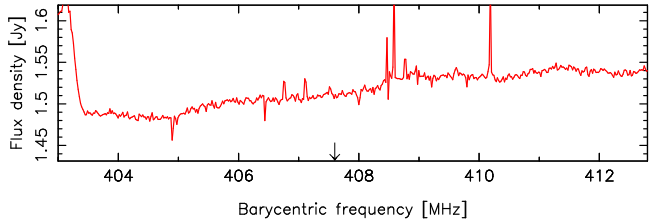}
\caption{GBT 408 MHz band spectrum towards 1402+044 hanning smoothed to a velocity resolution of 14 \kms.}
\label{1402_408}
    \end{figure}
2007 for a total of 7.2 hours, of which 6.1 hours were retained
($T_{\rm sys}\approx70$ K). The channel width was 9.766 kHz, giving a
spectral resolution of 7.2 \kms\ at 408 MHz (Fig. \ref{1402_408}).

1418--064 was observed using the spectral processor, tuned to a
central frequency of 320 MHz, on the 22nd of March 2007 for a total of
2.4 hours, of which 1.9 were retained ($T_{\rm sys}\approx80$ K). As
\begin{figure}
 \centering
\includegraphics[angle=270,scale=0.63]{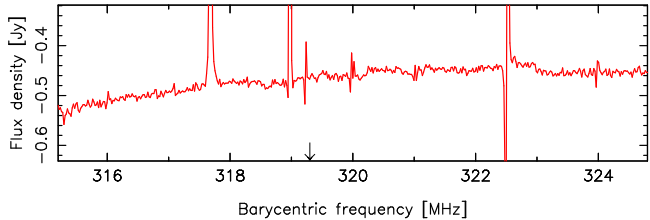}
\caption{GBT 320 MHz band spectrum towards 1418--064 hanning smoothed to a velocity resolution of 18 \kms.}
\label{1418-064}
    \end{figure}
per 0758+475, although the bandpasses were flat, the measured flux
density varied considerably between scans not permitting us to
determine a value for this (Fig. \ref{1418-064}). The spectral
resolution was 9.2~\kms\ at 319 MHz.

2136+141 was observed using the spectral processor, tuned to a
central frequency of 454 MHz, on the 5th and 7th
of January 2007. Originally, 12 hours were planned to search for 21-cm
absorption in the two (admittedly weak, Table~\ref{res}) absorbers towards this compact
\begin{figure}
 \centering
\includegraphics[angle=270,scale=0.63]{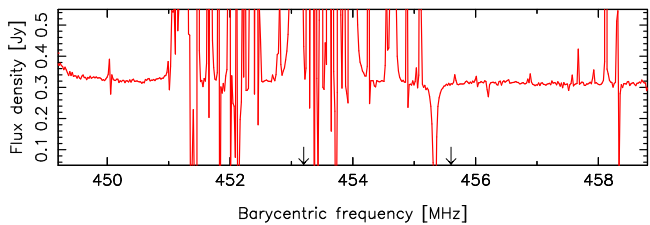}
\caption{GBT 454 MHz band spectrum towards 2136+141  hanning smoothed to a velocity resolution of 13 \kms.}
\label{2136+141}
    \end{figure}
quasar (Table \ref{size}), but due to
severe RFI over most of this band, we only observed for 2.0 hours
(where $T_{\rm sys}\approx65$ K). At redshifts of $z=2.118$ and $2.134$, the
10 MHz band centered on 454 MHz covered both 455.6 and 453.2 MHz at a
resolution of 6.5 \kms, with the RFI dominating 451 to 455 MHz.
This allowed us to only obtain a limit for the lower redshift sub-DLA, which
itself may not be reliable (Fig. \ref{2136+141}).

All of the data were reduced using the {\sc gbtidl} package. After
flagging, both polarisations were combined, linear (1st order)
baselines removed and the resulting data Gaussian smoothed, via the
{\sf gsmooth} task, to a resolution of $\geq3$ \kms. The results are
summarised in Table \ref{res}.

\subsection{GMRT 2008 observations}
\label{GMRT08}
\subsubsection{Motivation}

For the GMRT sample the intent was to find new high redshift ``end
points'' with which to verify the flattening off of $T_{\rm spin}/{f}$
at $\approx2000$~K at $z_{\rm abs}\gapp2$ \citep{ctp+07}. To this
end, we trawled the Sloan Digital Sky Survey Damped Ly{$\alpha$}
Survey Data Releases 1--3 \citep{phw05} for DLAs occulting radio-loud
quasars yet to be searched for 21-cm absorption. This yielded two DLAs
occulting quasars with flux densities in excess of $50$ mJy at
1.4~GHz, J080137.6+472528.2 and J081618.9+482328.4.

\subsubsection{Observations and data reduction}

For all of the observations, all 30 antennae were used with the 325 MHz
receiver, where for an antenna gain of 0.32 Ky Jy$^{-1}$ we expected
to reach an r.m.s. of $\sigma\approx3$ mJy per channel in each of the
two polarisations. For the backend we used a correlator bandwidth of
0.5 MHz, giving a velocity resolution of $\approx4$ \kms\ and a
coverage of $\approx\pm230$ \kms\ ($\Delta z\approx\pm0.003$).

J0801+472 (which is 0758+475, Sect. \ref{GBT06}) was observed at a
central frequency of 336.366 MHz on the 7th and 8th January
2008. 3C\,147 was used for the bandpass calibration and 0834+555 was
observed every 0.6 hours to correct for delays. In general, RFI was
minimal and the data were excellent with all 30 antennae operating,
although pairs involving antennae 13 \& 28 were flagged due to poor
phase stability. On the first night, the target was observed for a
total of 7.54 hours, of which only 0.14 hours required flagging and
350 good baseline pairs were retained.  On
\begin{figure}
 \centering
\includegraphics[angle=270,scale=0.63]{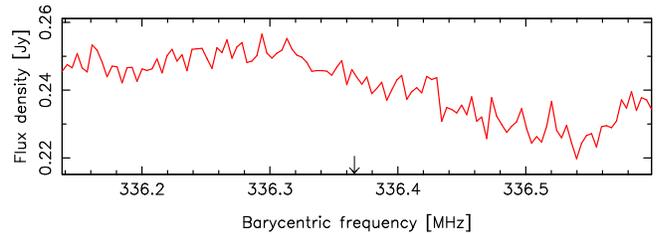}
\caption{GMRT 336 MHz band spectrum towards J0801+472 (0758+475). The velocity resolution is
3.5 \kms. }
\label{J0801+472}
    \end{figure}
the second night, the target was observed for a total of 7.78 hours of
which 7.64 hours were kept. Again the data were of high quality,
giving 400 good baseline pairs. After analysing each polarisation of
each night separately, a combined image was produced using the {\sf
  invert} task in {\sc miriad}. The source was unresolved by the
$13''\times11''$ beam.

J0816+4823 was observed at a central frequency of 320.214 MHz for a
total of 7.8 hours on 16th March 2008. 3C\,48 was used for the
bandpass calibration and again the phase calibrator 0834+555 was
observed every 0.6 hours during the observation.  After flagging the
non-operating antennae, 8 \& 28, as well as antenna 6, on which the phase
stability was poor, RFI appeared minimal and only a little time
dependent interference had to be removed (leaving 7.65 hours of
data). However, perhaps due to its low flux density, an accurate
image of the quasar could not be produced. We therefore flagged noisy
antenna pairs until the averaged visibilities gave an r.m.s. noise
level close to the $\approx10$ mJy expected.
\begin{figure}
 \centering
\includegraphics[angle=270,scale=0.63]{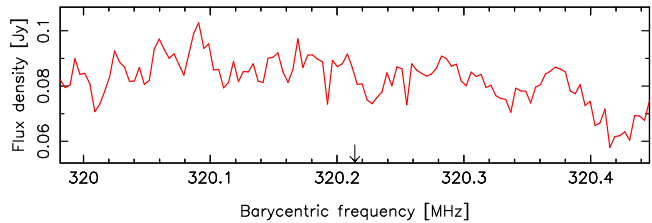}
\caption{GMRT 320 MHz band spectrum towards J0816+4823. The velocity resolution is
3.7 \kms. }
\label{J0816+4823}
    \end{figure}
Averaging the $\approx250$ remaining baselines in each of the
polarisations, gave a similar flux density to extracting this from the
cube, i.e. $85$ mJy at 0.32~MHz, which compares well with the $78$ mJy
at 1.4 GHz from the The Very Large Array's {\it Faint
  Images of the Radio Sky at Twenty Centimetres} (FIRST). Again, the
results are summarised in Table \ref{res}.

\subsection{GBT 2008 observations}
\label{GBT08}
\subsubsection{Motivation}

According to the 21-cm line strength--metallicity correlation ($T_{\rm
  spin}/f$\,--\,[M/H] anti-correlation, \citealt{ctp+07}), there
are three previously searched DLAs (PKS 0454+039, PKS 0528--250 \&
[HB89]~0957+561) for which the spin temperature/covering factor ratio
should be $T_{\rm spin}/{f}\lapp1000$ K, with the previous limits
possibly being close to those required to detect 21-cm absorption.  We
therefore applied for Green Bank Telescope time in order to improve
these limits to the point where 21-cm absorption would be detected,
 although a significant improvement was only managed for 0528--250
  (see Sects. \ref{or} \& \ref{lsmc}).

\subsubsection{Observations and data reduction}

As per the 2006 run, we used the GBT Prime Focus Receivers in position
switching mode, with the GBT spectrometer split into 32\,768 lags over a
12.5 MHz band, giving a channel spacing of 763~Hz, in dual
polarisation.

0454+039 was observed for a total of 2.9 hours over several runs from
28th August to 13th September 2008, at a central frequency of
763.8~MHz, giving a spectral resolution of 0.3 \kms. System
temperatures ranged from $T_{\rm sys}\approx30$ to 40~K over these
dates and the overall RFI was minimal, except over the frequency range
of interest, 763.6 to 763.8 MHz, where a blooming great spike appeared
in most of the spectra. This led to significant flagging, leaving 1.2
hours of
\begin{figure}
 \centering
\includegraphics[angle=270,scale=0.63]{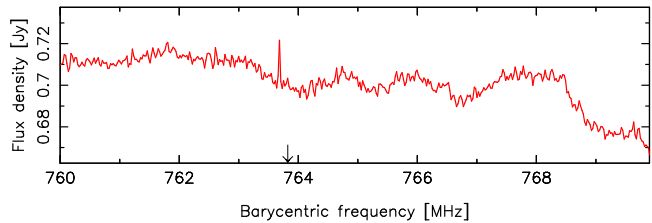}
\caption{GBT spectrum towards 0454+039, shown at a velocity resolution of 10 \kms.}
\label{0454+039}
    \end{figure}
good data, of which the average
spectrum is shown in Fig. \ref{0454+039}.

0528--250 was observed 2nd and 18th February 2008 for a total of 7.4
hours. At a central frequency of 327.7 MHz, giving a spectral
resolution of 0.7 \kms. The system temperature was typically $T_{\rm
  sys}\approx60$ K and RFI was only severe over 0.64 hours of the
observations. However, besides this there were intermittent RFI spikes
over the whole band, which only
\begin{figure}
 \centering
\includegraphics[angle=270,scale=0.63]{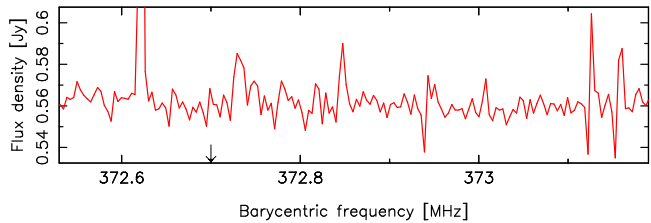}
\caption{GBT spectrum towards 0528--250 over the relatively RFI free 372.53--373.20 MHz, shown at a velocity resolution of 3 \kms.}
\label{0528-250}
    \end{figure}
permit r.m.s. estimates over the intervening clean regions. Therefore
the limit quoted in Table \ref{res} is for 372.53 to 373.20 MHz ($z =
2.8060-2.8116$), close to that expected from the redshift of the DLA (Fig. \ref{0528-250}).

\begin{figure}
 \centering
\includegraphics[angle=270,scale=0.63]{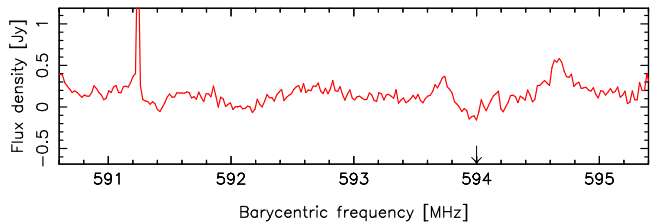}
\caption{GBT spectrum towards 0957+561A over the relatively RFI free 590.6--595.4 MHz, shown at a velocity resolution of 10 \kms.}
\label{0957+561}
    \end{figure}

0957+561A was observed for a total of 3.6 hours over several runs from
7th August to 8th November 2008, at a central frequency of 594 MHz,
giving a spectral resolution of 0.4 \kms. System temperatures ranged
from $T_{\rm sys}\approx30$ to 60 K over these dates and RFI was
severe. This led to extensive flagging of bad data, leaving only 14
minutes of ``good data'', of which the average spectrum is shown in
Fig. \ref{0957+561}.
Although this flagging removed all of the spikes close to 594 MHz, a strong ringing
was still present in the unsmoothed data and, as seen from the spectrum, no reasonable
limit can be obtained from these observations\footnote{Note that the unflagged data
give a flux density of $\approx1$ Jy, although the spectrum is plagued by RFI spikes close to the target frequency.}.

\section{Results}
\subsection{Observational results}
\label{or}

Typically the illumination of the optical (Lyman-\AL) and radio (21-cm)
absorbers arise from quite different sources sizes, these being $\lapp1$ pc
for the former, but in excess of 100 pc for the latter\footnote{Even
  over 10 kpc in some cases (see figure 4 of \citealt{cmp+03}).}
(\citealt{wgp05} and references therein). Thus, it is possible that 
the optical and radio observations could trace different
sight-lines (as is probably the case for 1622+238, \citealt{ctm+07}).
However, if the 21-cm and Lyman-$\alpha$ absorption do arise in the
same cloud complexes \citep{dl90}, the column density [\scm] of the
absorbing gas in a homogeneous cloud is related to the velocity
integrated optical depth, where
$\tau\equiv-\ln\left(1-\frac{\sigma}{f\,S}\right)$, of the 21-cm line
via \citep{wb75}:
\begin{equation}
N_{\rm HI}=1.823\times10^{18}\,T_{\rm spin}\int\!\tau\,dv\,,
\label{enew}
\end{equation}
where $T_{\rm spin}$ [K] is the mean harmonic spin temperature of the gas, $\sigma$
is the depth of the line (or r.m.s. noise in the case of a
non-detection) and $S$ \& $f$ the flux density and covering factor of
the background continuum source, respectively. In the optically thin
regime ($\sigma\ll f.S$), Equ. \ref{enew} reduces to $N_{\rm
  HI}=1.823\times10^{18}\frac{T_{\rm spin}}{f}\int\!\frac
{\sigma}{S}\,dv$, which we apply to the entries in Table \ref{res},
where we summarise the results for the three sets of observations.
\begin{table*}
\centering
\begin{minipage}{165mm} 
\caption{Our search results. $z_{\rm abs}$ is the redshift of the DLA
  giving the observed 21-cm frequency, $\nu_{\rm obs}$. {\sc run} gives the
  observing run (Sects. \ref{GBT06} to \ref{GBT08}), $S$ is the flux
  density at $\nu_{\rm obs}$, $\sigma$ is the r.m.s. noise level reached per
  $\Delta v$ channel [\kms],
  where these are $>3$ \kms; for the GBT observations the resolutions
  are typically much finer (Sect. \ref{GBT06}). Since the measured
  r.m.s. noise is dependent upon the spectral resolution, as usual
  \citep{cmp+03,ctp+07}, for the non-detections the $3\sigma$ upper
  limits of $\tau_{\rm peak}$ at a velocity resolution of 3 \kms\ are
  quoted. The total neutral hydrogen column density, $N_{\rm HI}$
  [\scm], is given with the corresponding reference, which yields the
  quoted spin temperature/covering factor ratio, via the estimated
  FWHM [\kms] for the non-detections (see main text).\label{res}}
\begin{tabular}{@{}l  c c c c l c  c c c c r@{}} 
\hline
QSO & $z_{\rm abs}$ &$\nu_{\rm obs}$ [MHz] & {\sc run} & $S$ [Jy] & $\sigma$ [mJy] & $\Delta v$  &  $\tau_{\rm peak}$ &$\log_{10}N_{\rm HI}$ & Ref. & FWHM & $T_{\rm spin}/f$ [K]\\
\hline
0454+039  &  0.8596 &  763.8   &GBT08& 0.70    &  $<4.6$    &  3.0   & $<0.020$  & 20.7 & S95 & 19 & $>690$\\ 
0528--250 &  2.811  & 372.7   &GBT08&  0.561    & $<6.9^a$   &  3.0   & $<0.037$  & 21.3 & L96 &40  & $>700$\\ 
J0801+4725$^{*}$ & 3.2228 & 336.4   &GBT06 & --- & $<6.1$  & 3.0  & ---   & 20.7 & P05  & --- &---\\
 ...       &  ...   &...    & GMRT & 0.24 & $<2.9$ &3.5 &  $<0.039$ & ... & ... & 20$^{\dagger}$ & $>330$ \\ 
J0816+4823 & 3.4358 & 320.2  & GMRT & 0.08  &$<6.3$ & 3.7 & $<0.29$ & 20.8 & P05 & 20$^{\dagger}$&  $>60$\\
0957+561A  & 1.391  & 594.0  & GBT08&  --- &  ---  & ---  & ---      & 20.3 & T93    & 30 & --- \\ 
1402+044   &  2.713 & 382.5  & GBT06& 1.41 & $<16$ & 3.0  & $<0.034$ & ---   & T89 &--- & --- \\ 
...        &  2.708 & 383.1  & GBT06& 1.41 & $<16$ & 3.0  & $<0.034^b$  & 20.9 & P04 & 20$^{\dagger}$&$>610$ \\ 
...        &  2.688 & 385.1  & GBT06& 1.37 & $<20^c$ & 3.0&  $<0.044$   & ---   & T89 & ---& --- \\
...        &  2.485 & 407.6  & GBT06& 1.51 & $<4.8$ & 7.2 &$< 0.015$ &   20.2 &T89  & 20$^{\dagger}$ & $>270$ \\
1418--064   &  3.449 & 319.3  & GBT06&$0.374^d$ &$<8.1$&9.2& $<0.12$  &   20.4 & E01 &25 &  $>40$\\ 
2136+141   &  2.134 & 453.2  & GBT06& ---   &   ---     & ---    &   ---   & 19.8 & T89 &--- & ---\\ 
...        &  2.118 & 455.6  & GBT06& 0.320& $<7.7$ &  6.4& $<0.11$ & 19.8 & T89 &20$^{\dagger}$ & $>10$ \\    
\hline
\end{tabular}
{Notes: $^{*}$0758+475 -- observed with both the GBT and GMRT
  (Sects. \ref{GBT06} \& \ref{GMRT08}). $^{\dagger}$No Mg{\sc \,ii}
  equivalent width or metallicities found, so $20$
  \kms\ assumed. $^a$Over 372.53 to 373.20 MHz ($z =
  2.8060-2.8132$). $^b$Double the sensitivity reached with the WSRT
  observations \citep{ctp+07}, so we adopt this limit. $^c$Subject to
  an RFI spike from 385.17 to 385.23 MHz ($z = 2.6872-2.6877$).
  $^d$Flux density could not be determined from our observations and
  so have adopted value for TXS 1418--064 at 0.374 Jy \citep{dbb+96},
  although this is $4'6''$ from PKS B1418-064 (for which
  $S_{\rm 2.7 GHz} = 0.38$ Jy \& $S_{\rm 5.0 GHz} = 0.34$ Jy) it is
within the Texas survey beam.\\ References: T89 --
  \citet{twl+89}, T93 -- \citet{tb93}, S95 -- \citet{sbbd95}, L96 -- \citet{lsb+96}, E01 -- \citet{eyh+01}, P04 -- \citet{ph04}, P05 --
  \citet{phw05}.}
\end{minipage}
\end{table*}

For our and the previous non-detections we have no knowledge of what
the FWHM of the 21-cm profile should be, thus hampering the
determination of a lower limit for $T_{\rm spin}/f$. Previously
\citep{cmp+03,ctp+07}, we have used the $\approx20$ \kms\ mean of the
detections to derive velocity integrated optical depth limits for the
non-detections. An alternative to this is to estimate the FWHMs from
the spin temperature, via the kinetic temperature according to $T_{\rm
  spin}\approx T_{\rm kin}\lapp22\times{\rm FWHM}^2$
(e.g. \citealt{kc02}), although in the absence of detailed knowledge
of the covering factor, it is impossible to assign a specific spin
temperature to any of the DLAs. In light of the findings of
\citet{ctp+07}, however, we can make educated guesses to the profile
widths according to the FWHM-- Mg{\sc \,ii} rest-frame equivalent
width correlation for the detections. This gives
FWHM\,$\approx13\,{\rm W}_{\rm r}^{\lambda2796}$ (figure 6 of
\citealt{ctp+07}), and, where the Mg{\sc \,ii} equivalent widths are
not available (generally at $z_{\rm abs}\gapp2.2$), we estimate these
from the metallicity via ${\rm W}_{\rm
  r}^{\lambda2796}\approx2.0\,{\rm [M/H]} + 4.0$ (figure 7 of
\citealt{ctp+07}, see also \citealt{mcw+07})\footnote{Applying the
  above conversions gives a range of $5 - 42$ \kms\ for the
  non-detections, thus not drastically changing the values of
  $\log_{10}(T_{\rm spin}/f)$, shown in the following figures, from
  those using a FWHM of 20 \kms.}. Note that, although
\citet{gsp+09a,kpec09} find that the FWHM--${\rm W}_{\rm
  r}^{\lambda2796}$ correlation does not generally hold for Mg{\sc
  \,ii} absorption systems, when only those which are likely to be
DLAs (i.e. with an Mg{\sc \,i} 2852 \AA\ equivalent width of ${\rm
  W}_{\rm r}^{\lambda2852}>0.5$ \AA, \citealt{rtn05}) are added to the
sample, the correlation is still found to hold for all suspected DLAs
\citep{cur09a}.

Commenting on our observational results, due to the effects of RFI and
weak radio fluxes, we cannot assign limits to four of the absorbers
and the limits for a further three are so poor as to be of little
use. Of the remaining five, for 0454+039 we have matched the previous
limit of $T_{\rm spin}/f\gapp600$~K \citep{bw83}, while for 0528--250
we have significantly improved upon the $T_{\rm spin}/f\gapp100$~K
limit of \citet{cld+96}\footnote{\label{cld}Based upon our 3
  \kms\ resolution and 40 \kms\ line-width (Table~\ref{res}).}. This
is the first search for 21-cm absorption at $z_{\rm abs}=3.223$
towards J0801+4725, although, due to the relatively low flux density,
the limit is only $T_{\rm spin}/f\gapp300$~K.\footnote{For this
    and the other GMRT target, J0816+4823, it is possible that the
    redshift coverage of $z_{\rm abs}\pm0.003$ is not sufficient to
    cover uncertainties in the absorber redshift, which are determined
    from the Lyman-\AL\ line \citep{phw05}. However, J0801+4725 has
    been searched over $z_{\rm abs}=3.30\pm0.08$ with the GBT
    (Fig.~\ref{0758+475}) and the J0816+4823 $T_{\rm spin}/f$ limit is
    too poor in any case to contribute to the analysis.} For the $z_{\rm
  abs}=2.708$ DLA towards 1402+044, we have improved the previous
limit \citep{ctp+07} by a factor of two and for the $z_{\rm
  abs}=2.485$ sub-DLA, publish the first limit, although this is weak
due to the low neutral hydrogen column density.

\subsection{Update of the previous results}
\label{oftpr}

\begin{figure*}
\centering
\includegraphics[angle=270,scale=0.73]{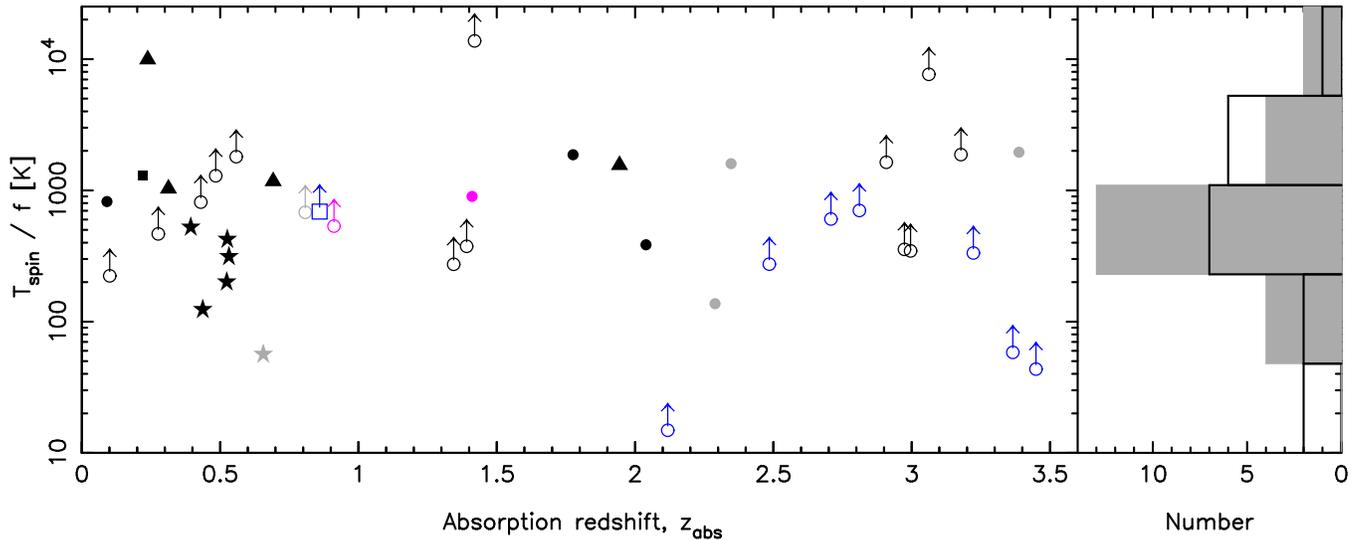}
\caption{The spin temperature/covering factor ratio versus the
  absorption redshift for the DLAs searched in 21-cm absorption. The shapes represent the type of galaxy with which
  the DLA is associated: circle--unknown type, star--spiral,
  square--dwarf, triangle--LSB. The filled symbols represent the 21-cm
  detections and the unfilled symbols+arrows show the lower limits for
  the non-detections (all of these bar one, 0454+039 at $z_{\rm
    abs}=0.8596$, have unknown host identifications). The  grey symbols
  show the results since \citet{cmp+03} and the coloured symbols
  show the new results presented here [blue] as well as those of \citet{kpec09} [magenta].
The solid grey histogram shows the spin temperature/covering factor distribution for
the low redshift ($z_{\rm abs}<1.6$) absorbers and the outlined histogram the distribution for
the high redshift ($z_{\rm abs}\geq1.6$) absorbers, where the limits to $T_{\rm spin}/f$
are treated as detections.}
\label{Toverf}
\end{figure*}

Adding the new results and applying the estimated FWHMs, in Fig.~\ref{Toverf} we update
the spin temperature/covering factor--absorption redshift distribution of \citet{cmp+03}.
From the histogram it is seen that the low redshift absorbers have a
similar $T_{\rm spin}/f$ distribution to those at high redshift
($z_{\rm abs}\geq1.6$) and in Table~\ref{stats} we present the
results of the statistical tests performed on the low and high
redshift samples.
\begin{table*}
\centering
\begin{minipage}{150mm} 
\caption{The statistics for and between the low and high redshift
  samples for various redshift cuts, where $P$ gives the probability
  that the low and high redshift DLAs are drawn from the same sample and $S$ gives the corresponding
  significance (assuming Gaussian statistics). The values in the first
  four rows incorporate the limits to the non-detections via the {\sc
    asurv} package \citep{ifn86} and for the Kolmogorov-Smirnov test
  the limits are treated as detections (as per \citealt{kc02}).
\label{stats}}
\begin{tabular}{@{}l  c c c c ccc@{}} 
\hline
& \multicolumn{2}{c}{$z_{_{\rm CUT}} = 1.0$} & \multicolumn{2}{c}{$z_{_{\rm CUT}} = 1.6$} & \multicolumn{2}{c}{$z_{_{\rm CUT}} = 2.0$}\\
 & $z_{\rm abs } < z_{_{\rm CUT}} $ & $z_{\rm abs }\geq z_{_{\rm CUT}}$ & $z_{\rm abs } < z_{_{\rm CUT}} $ & $z_{\rm abs }\geq z_{_{\rm CUT}}$ & $z_{\rm abs } < z_{_{\rm CUT}} $ & $z_{\rm abs }\geq z_{_{\rm CUT}}$\\
\hline
Sample size, $n$ &  19 & 20& 23& 18 &25 & 16\\
Mean, $\overline{\log_{10}(T_{\rm spin}/f)}$&$3.08\pm0.18$  &$3.38\pm0.18$ &  $3.16\pm0.17$ & $3.26\pm0.16$ &  $3.11\pm0.15$ & $3.30\pm0.20$\\
$P_{\rm Log-rank}$ & \multicolumn{2}{c}{0.0718} &  \multicolumn{2}{c}{0.265} & \multicolumn{2}{c}{0.203}\\
$S_{\rm Log-rank}$ & \multicolumn{2}{c}{$1.80\sigma$} & \multicolumn{2}{c}{$1.39\sigma$}  & \multicolumn{2}{c}{$1.27\sigma$}\\
$P_{\rm KS}$ &  \multicolumn{2}{c}{0.443} &  \multicolumn{2}{c}{0.448} &  \multicolumn{2}{c}{0.358}\\
$S_{\rm KS}$ &\multicolumn{2}{c}{$0.767\sigma$} &  \multicolumn{2}{c}{$0.759\sigma$} &  \multicolumn{2}{c}{$0.919\sigma$} \\
\hline
\end{tabular}
\end{minipage}
\end{table*}
These were done for redshift cuts of $z_{_{\rm CUT}} = 1.0$, $1.6$
(the turnover in angular diameter distance, Sect. \ref{copr}) and
$2.0$ (as \citealt{kc02}).  From these, we find slightly lower
  mean values of $T_{\rm spin}/f$ for the low redshift absorbers, with
  log-rank and Kolmogorov-Smirnov tests indicating that there is
  little significance in the differences between the low and high
  redshift samples. Bear in mind, however, that the FWHMs for the
  non-detections have been estimated (Sect. \ref{or}), with some of
  these relying on a secondary correlation (FWHM\,$\propto\,{\rm
    W}_{\rm r}^{\lambda2796}\approx2.0\,{\rm [M/H]} + 4.0$), when the
  Mg{\sc \,ii} equivalent width is not available\footnote{Setting
    the FWHMs obtained from the metallicity to 20 \kms, however,
    alters only the $z_{\rm abs }\geq z_{_{\rm CUT}}$ averages in
    Table \ref{stats}, and these only slightly. Overall, there is very
    little change with $S_{\rm Log-rank} = 1.95\,,1.45$ \&
    $1.45\sigma$ and $S_{\rm KS} = 1.15\,,1.22$ \& $0.42\sigma$,
    respectively.}. Furthermore, the Kolmogorov-Smirnov test treats
  the upper limits as actual values and higher sensitivity limits at
  high redshift are required in order to confirm the similarities
  between the samples apparent in Table \ref{stats}.

Performing a Kolmogorov-Smirnov test on the redshift
  distributions between the detections and non-detections, thus
  requiring no assumptions, gives an 11\% probability (significant at
  $1.6\sigma$) that these are drawn from the same
  sample. Additionally, the non-detections have a higher mean redshift
  -- $\overline{z_{\rm abs}} = 1.77$ ($\sigma=1.08$), compared to
  $\overline{z_{\rm abs}} = 1.13$ ($\sigma=0.96$) for the detections.
  Therefore, we cannot rule out that the $T_{\rm spin}/{f}$
  distribution differs between the two redshift regimes. Although the
  relative contributions of $T_{\rm spin}$ and ${f}$ are unknown, the
  fact that the largest differences are seen for $z_{_{\rm CUT}} =
  1.0$ (where $\overline{[T_{\rm spin}/{f}]_{z < 1}}=1200$~K cf.
  $\overline{[T_{\rm spin}/{f}]_{z \geq 1}}=2400$~K) is
  consistent with geometry effects giving larger values of $f$ at
  $z_{\rm abs} <1$ (\citealt{cw06} and Sect. \ref{copr}).

While we are discussing statistical trends over the whole sample,
it should be emphasised that the 21-cm absorption searches yield a
very heterogeneous sample, with each survey having different selection
criteria (including the three here), resulting in a mix of detections
and non-detections across various absorber morphologies (where these
are known, Fig. \ref{Toverf}). However, presuming that the optical
and radio absorption clouds are coincident, Equ. \ref{enew} accounts
for differences in the neutral hydrogen column densities and the flux
densities of the background sources (discussed in detail in
\citealt{cmp+03}), leaving only the $T_{\rm spin}/f$ degeneracy.
Naturally, various properties of the absorber and emitter, as well as the
distance between them, could influence this ratio, issues which we
discuss here.  However, whether it be due to similar values of $T_{\rm
  spin}$ {\em and} $f$ for all of the absorbers, or just the ratio being
maintained, there may be no large differences between the low and high
redshift samples. Furthermore, since $f\leq1$, the ordinate in
Fig.~\ref{Toverf} represents the maximum possible spin temperatures
for the detections of which there are now several at $z_{\rm
  abs}\gapp2$. This indicates that high redshifts do not {\em necessarily}
entail high spin temperatures, as advocated by \citet{kc02} [although
a larger fraction of warm gas \citep{klm+09} does not exclude the
possibility of cold clumps]. Case in points against exclusively high
spin temperatures at high redshift are,
\begin{enumerate}
\item The 21-cm  detection at $z_{\rm abs}=2.289$ towards
0311+430 \citep{ykep07}, which has one of the lowest ``spin
temperatures'' yet found, i.e. $T_{\rm spin}/(f=1)\approx140$~K.
\item The excitation temperature of $T_{\rm ex}\approx 50$~K 
derived from the \MOLH\ transitions in the $z_{\rm abs}=2.626$ DLA
towards 0812+32B, indicating the presence of cold gas \citep{jwpc09}.
\item The other 17 DLAs in which \MOLH\ has been detected
(\citealt{nlps08} and references therein). These are generally
at $z_{\rm abs}\gapp1.7$, although they may represent a specific
class of absorber within the general DLA population \citep{cwmc03,mcw04}.
\end{enumerate}

Lastly, \citet{cmp+03} also noted a correlation between the integrated 21-cm
optical depth and the \HI\ column density. 
\begin{figure}
\vspace{6.5cm}
\includegraphics{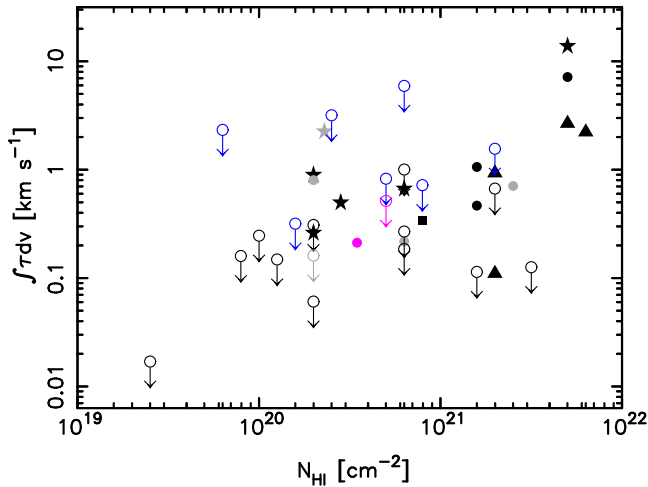}
\caption{The velocity integrated optical depth of the 21-cm absorption
versus the total neutral hydrogen column density. The
  symbols are as per Fig. \ref{Toverf}.}
\label{tau-N}
\end{figure}
Updating this for the DLAs (Fig. \ref{tau-N}) and incorporating the
upper limits (as above), there is a
probability of $P(\tau)=0.0021$ that the correlation occurs by chance.
This corresponds to a $3.08\sigma$ significance, and since the gradient 
is $\propto f/T_{\rm spin}$, suggests that, on
the whole, the ratio of spin temperature to covering factor is a
constant for DLAs, although there is some considerable scatter.

\section{Discussion}
\subsection{Covering factors}
\subsubsection{Angular diameter distances and other effects}


\label{copr}

As mentioned in Sect. \ref{GBT06}, \citet{cmp+03} suggested that,
contrary to the high redshift equals high spin temperature argument of
\citet{kc02}, 21-cm absorption should be readily detectable at high
redshift towards compact radio sources, where the covering factor is
maximised. 
At the time of writing up the results presented in
\citet{cw06}, this was confirmed by the detection of 21-cm absorption
in the highest redshift example to date, at $z_{\rm abs}=2.347$
towards 0438--436 (\citealt{kse+06}), which has a background radio
source size of only $0.039''$ at 5 GHz \citep{tml+98}. Following this,
since submitting the proposal for the 2006 GBT observations, 21-cm
absorption has been confirmed at yet higher redshift, $z_{\rm
  abs}=3.386$ towards 0201+113 (\citealt{kcl06}, see also
\citealt{dob96,bbw97}), as well as at $z_{\rm abs}=2.289$ towards
0311+430 \citep{ykep07}. Again, both of these occult compact radio
sources of $<0.007''$ at 1.4 GHz \citep{sbom90}\footnote{\citet{kcl06}
  report a de-convolved source size of $17.6\times6.6$ mas$^2$ at 328
  MHz.} and $1.36''\times0.63''$, respectively. The latter was noted
by \citet{ykep07} to be unusual in its low spin temperature of
$T_{\rm spin}\leq140$~K at such a high redshift, although this is
not unusual in regard to the finding of \citet{cmp+03,cw06}.

From Table \ref{size} we see that five of our targets are towards
radio source sizes of $\theta_{\rm QSO}\lapp1''$ (recently found
  to be $0.07''$ for 0454+039 and $0.17''$ for 0528--250,
  \citealt{klm+09}), although we have only meaningful $T_{\rm spin}/f$ limits for three
of these (0454+039, 0528--250 \& 0758+475 --
Table~\ref{res}). However, all of the GBT 2006 targets
(Sect. \ref{GBT06}) were selected before the arguments of \citet{cw06}
were formulated, and so all of the 2006 targets (and two of the three
just mentioned), are all at high redshift, and will therefore be
disadvantaged with respect to the DLA-to-QSO angular diameter distance
ratios.

\begin{figure*}
\centering
\includegraphics[angle=270,scale=0.73]{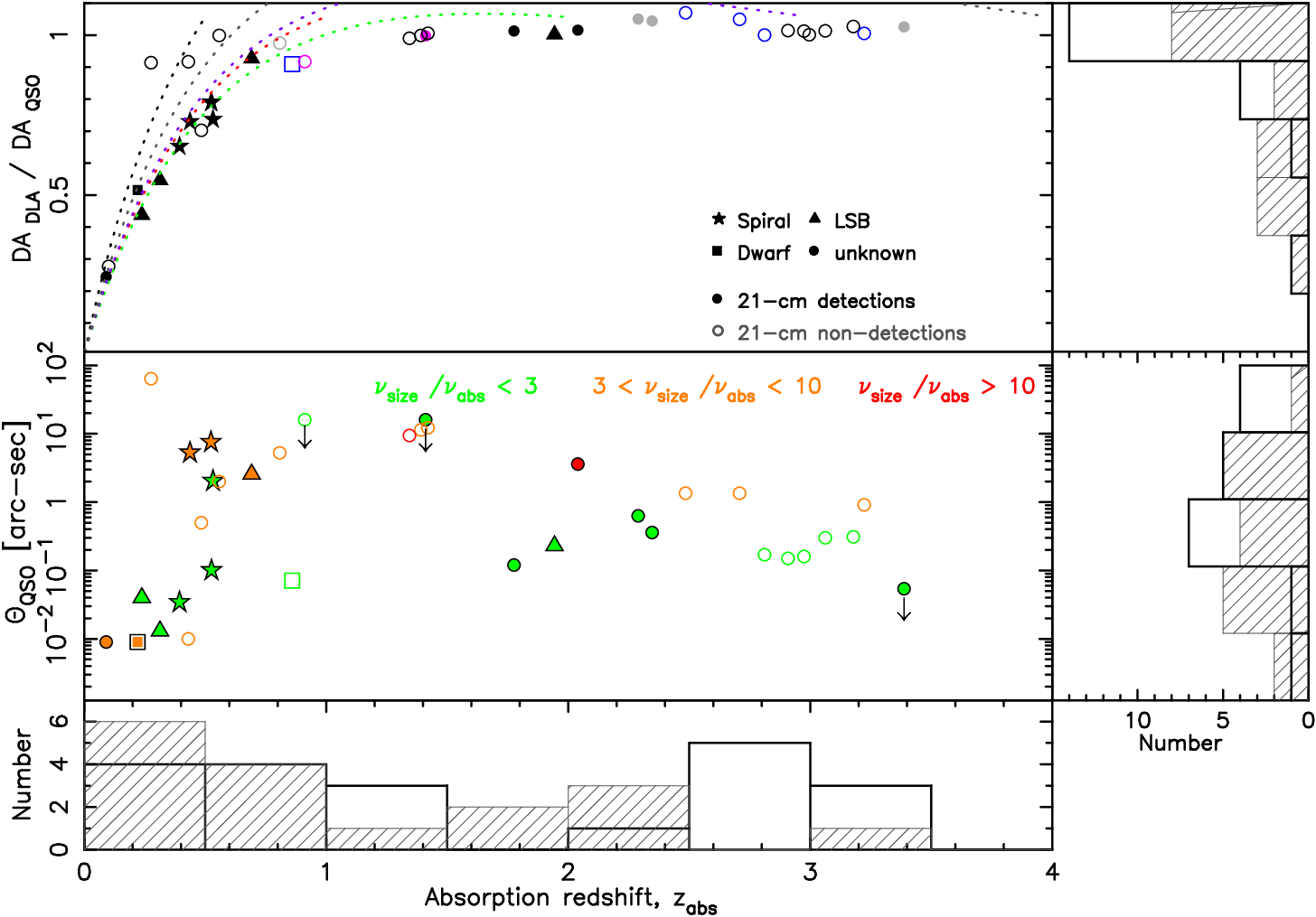}
\caption{Top: The absorber/quasar angular diameter distance ratio
  versus the absorption redshift. Updated from \citet{cw06}, where the
  symbols are as per Fig. \ref{Toverf} and we include only the DLAs
  which have been searched to $T_{\rm spin}/f\geq100$~K (see main
  text). The iso-redshift curves show how $DA_{\rm DLA}/DA_{\rm QSO}$
  varies with absorption redshift, where $DA_{\rm QSO}$ is for a given
  QSO redshift, given by the terminating value of $z_{\rm abs}$
  (throughout this paper we use $H_{0}=71$~km~s$^{-1}$~Mpc$^{-1}$,
  $\Omega_{\rm matter}=0.27$ and $\Omega_{\Lambda}=0.73$). That is, we
  show the $DA_{\rm DLA}/DA_{\rm QSO}$ curves for $z_{\rm em}$ = 0.5,
  1, 2, 3 \& 4. 
  Bottom: The radio source size versus the absorption
  redshift. The symbols are colour coded according to the ratio of the
  frequency of the image to the redshifted 1420 MHz frequency: green
  -- $\nu_{\rm size}/ \nu_{\rm abs}\leq3$, orange -- $3<\nu_{\rm
    size}/ \nu_{\rm abs}\leq10$ and red -- $\nu_{\rm size}/ \nu_{\rm
    abs} > 10$. The hatched histogram represents the detections and the
bold histogram the non-detections.}
\label{2-distance-z}
\end{figure*}
Showing these in Fig. \ref{2-distance-z} (top), we see that all but
one of our target DLAs (for which we have limits), are indeed at
angular diameter distance ratios of $DA_{\rm DLA}/DA_{\rm
  QSO}\approx1$, due to their high redshift selection. For a given
absorber and radio source size, this will disadvantage the effective
covering factor (possibly giving the observed
  $\overline{[T_{\rm spin}/{f}]_{z \geq 1}}\approx2\,\overline{[T_{\rm
        spin}/{f}]_{z < 1}}$, Sect. \ref{oftpr}), although, as
mentioned above, the three new high redshift detections occult compact
radio sources, as shown in the bottom panel of
Fig. \ref{2-distance-z}, where we use the source sizes compiled in
\citet{cmp+03} updated with those given in
\citet{klm+09}\footnote{\label{foot7}Many of the source sizes are
  measured at (VLBI) frequencies, which are often many times higher
  than that of the redshifted 21-cm line (see table 2 of
  \citealt{cmp+03}), and so in Fig. \ref{2-distance-z} (bottom) we
  have flagged each of the DLAs according to how many times larger
  than $1420/(z_{\rm abs}+1)$ the frequency at which the source size
  is measured, i.e. $\nu_{\rm size}/\nu_{\rm abs}$. The recently
  published low frequency VLBA imaging of \citet{klm+09} accounts for
  many of the ``green'' values, where $\nu_{\rm size}/ \nu_{\rm
    abs}\leq3$. Since a number of these sources appear resolved, we
  have recalculated the extents, rather than using the deconvolved
  sizes quoted in \citet{klm+09}, which are often smaller than the
  beam and may be unphysical (e.g. $\sim0$ pc for
  0405--331).}. However, all of our non-detections (which are
predominately at $z_{\rm abs}\gapp2.4$), also occult similar radio
source sizes. Of these, three of the sizes are not as reliable (orange
symbols where $3 < \nu_{\rm size}/ \nu_{\rm abs}<10$, cf. green where
$\nu_{\rm size}/ \nu_{\rm abs}\leq3$), although as seen from
Fig. \ref{Toverf}, only three published non-detections are
sufficiently sensitive to detect $T_{\rm spin}/{f}\gapp2000$~K at
$z_{\rm abs}\gapp2$ (Sect. \ref{GMRT08}).

While, due to geometry effects, a high angular diameter distance will
disadvantage the absorption cross-section in comparison to the same absorber placed
at a low angular diameter distance, with the 21-cm searches spanning a
look-back time range of 12 Gyr there are also evolutionary effects to
be considered. For instance, an evolution in the morphologies of the
absorbing galaxies is expected to result in different metallicities
and these heavy elements can provide cooling pathways for the diffuse
gas (\citealt{dm72}). Furthermore, it is shown that the relative mix
of the cold neutral medium (CNM, where $T\sim150$ K and $n\sim10$
\ccm) and the warm neutral medium (WNM, $T\sim8000$ K and $n\sim0.2$
\ccm) can be affected by the metallicity \citep{whm+95}.

At low redshifts, where the DLA hosts can be imaged directly, the
absorbing galaxies appear to be a mix of spirals, dwarf and LSBs
(e.g. \citealt{lbbd97,rnt+03,cl03}) and, in general, large galaxies
appear to be more populous at low redshift with smaller morphologies
becoming more common at high redshifts \citep{bmce00,lf03}.
Therefore, in the more compact absorbing galaxies, where the
metallicities are lower than in the larger spirals (see
Sect. \ref{lsmc}), we may expect the gas to be more dominated by the
WNM and, as shown by \citet{ctp+07}, $T_{\rm spin}/{f}$ is
anti-correlated with the metallicity (see Sect. \ref{lsmc}). However, although the
metallicity is itself anti-correlated with the look-back time
(\citealt{pgw+03,rphw05}), no [M/H]--$z_{\rm abs}$ correlation is seen
for the 21-cm searched sample \citep{ctp+07}, and, as seen in 
Fig.~\ref{2-distance-z}, very few morphologies are known for these at
$z_{\rm abs}\gapp1$. Furthermore, the $T_{\rm spin}/{f}$\,--\,[M/H]
anti-correlation is consistent with {\em either} the spin temperature
being lower or the covering factor being larger at high metallicities,
with the larger galaxies likely to have high values of $f$ as well as
[M/H]. Although the $T_{\rm spin}/{f}$ degeneracy is unbreakable, it
is possible that both factors contribute and that the values of [M/H],
$T_{\rm spin}$ and ${f}$ are in fact interwoven \citep{ctp+07}.

In addition to decreasing metallicities, the increase in the
background ultra-violet flux will increase the ionisation fraction of
the gas \citep{bdo88}, in addition to raising its spin temperature
\citep{fie59,be69}, an effect also seen in associated systems, where
21-cm absorption has never yet been detected in the host galaxy of a
quasar with $L_{\rm UV}\gapp10^{23}$ W Hz$^{-1}$
\citep{cww+08}. Therefore, as well as always having large angular
diameter distance ratios and lower metallicities, high redshift
absorption systems will generally be subject to higher ultra-violet fluxes,
although for intervening absorbers the effect will not be as severe as
for associated absorbers. Nevertheless, this could be
the cause of many of the $T_{\rm spin}/{f}$ lower limits at $z_{\rm
  abs}\gapp2.5$ (Fig. \ref{Toverf}), although, again, such heated gas
could have a significantly lower 21-cm absorption cross-section as well as
a higher temperature.

\subsubsection{Absorber extents}
\label{aex}

\begin{figure}
\vspace{3.0cm}
\includegraphics{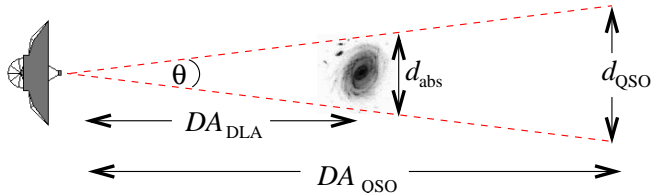}
\caption{In the small angle approximation $\theta = d_{\rm abs}/DA_{\rm DLA} = d_{\rm QSO}/DA_{\rm QSO}$, 
giving $d_{\rm abs (min)}=\theta_{\rm QSO}.\,DA_{\rm DLA}$ when the absorber just covers the
  background radio source size.}
\label{dla_schem}
\end{figure}

\begin{figure}
\vspace{6.5cm}
\includegraphics{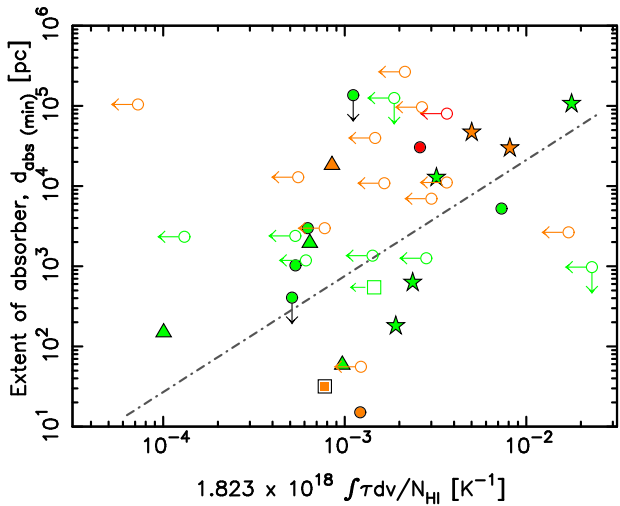}
\caption{The minimum extent of the absorber required to cover the radio source
  versus the normalised line strength. The symbols are as per
  Fig. \ref{2-distance-z} (bottom). The line shows the least-squares fit to the
  detections. The upper limits in the
  ordinate are due to upper limits in the background radio source
  sizes.}
\label{size-z}
\end{figure}

In order to consider the radio source size in conjunction with the
ratio of the angular diameter distances, we calculate the extent of
absorber required to fully cover the radio source
(Fig.~\ref{dla_schem}) and show this against the normalised line
strength, $1.823\times10^{18}\,\int\tau\,dv/N_{\rm HI}= f/T_{\rm spin}$ (Fig. \ref{size-z}).
From this we see a correlation between the extent of the absorber and
the 21-cm absorption strength for the detections. For the 18
detections this is significant at $2.7\sigma$, falling to $2.2\sigma$
for the 12 detections with $\nu_{\rm size}/ \nu_{\rm abs}\leq3$ (``green''), with
the non-detections exhibiting no correlation whatsoever (bringing the
significance for the whole 39 strong sample down to $1.4\sigma$)\footnote{Using the
deconvolved values of \citet{klm+09}, rather than those measured (see footnote \ref{foot7}),
gives $2.5\,,1.9$ \& $1.3\sigma$, respectively.}.

The fit in Fig. \ref{size-z} suggests that the non-detections which
have been searched sufficiently deeply, (i.e. neglecting the two with
$f/T_{\rm spin}\gapp0.01$), may require large absorbers ($\gapp10$~kpc
scales) in order to achieve effective covering factors. As is seen
from the figure, several detections also occupy this regime, although
these tend to be spiral galaxies.  Therefore, like \citet{kc02}, and
as discussed in \citet{cmp+03}, the majority of non-detections (which
are at high redshift and have unknown morphologies) may be
non-spirals.  Although, unlike \citet{kc02}, we again suggest that may
be their small sizes, in addition to/rather than high mean spin
temperatures, which may be responsible for their weak line strengths.

The covering factor can be defined\footnote{It can also be
estimated as the ratio of the compact unresolved component's
flux to the total radio flux \citep{bw83,klm+09}. However, even if
high resolution radio images at the redshifted 21-cm frequencies are
available, this method gives no information on the extent of the
absorber (or how well it covers the emission). Furthermore, the high
angular resolution images are of the continuum only and so do not give
any information about the depth of the line when the extended
continuum emission is resolved out.} as $f\equiv [d_{\rm
  abs}/(\theta_{\rm QSO}.\,DA_{\rm DLA})]^2$ \citep{cw06}, meaning
that, in general\footnote{That is, where $d_{\rm abs}$ is the actual
  extent of the absorber, rather than the extent minimum required to cover
  $\theta_{\rm QSO}$.}, the ordinate in Fig. \ref{size-z} is
equivalent to $d_{\rm abs}/\sqrt{f}$.  This therefore suggests, at
least for the detections, that the line strength increases with the
size of the 21-cm absorption region, a trend which is apparent, even
when the absorber size is weighed down by the $\sqrt{f}$ factor.
and, in general, from the normalised line strength,
\begin{equation}
1.823\times10^{18}\,\frac{\int\tau\,dv}{N_{\rm HI}}=  \frac{f}{T_{\rm spin}} \propto\theta_{\rm QSO}.\,DA_{\rm DLA} = \frac{d_{\rm abs}}{\sqrt{f}}.
\end{equation}
That is, $d_{\rm abs}\propto f^{3/2}/T_{\rm spin}$, which makes sense
in terms of both covering factor and spin temperature -- a larger
absorber implies a larger covering factor, as well as the 
  possibility of a lower spin temperature. Presuming that this is not
dominated by the numerator, this {\em could} suggest that the larger
absorbing galaxies are subject to lower spin temperatures
(\citealt{kc02} and references therein).  However, one interpretation
of the distribution of the non-detections along the ordinate in
Fig. \ref{size-z} is that for a given absorption cross-section
(i.e. assuming the same $d_{\rm abs}$ for all the DLAs), the covering
factor is generally lower than for a number of the detections.

Note that, in addition to geometry effects, it is possible that these
lower covering factors could also be caused by lower beam filling
factors, introduced by structure (``clumpiness'') in the absorbing
gas, which may be behind the 21-cm variability seen at $z_{\rm
  abs}=0.313$ towards 1127--145 \citep{kc01} and at $z_{\rm
  abs}=3.386$ towards 0201+113 \citep{dob96,bbw97,kcl06}.
\citet{gsp+09a} also note from their large survey of Mg{\sc \,ii}
absorption systems that the decreasing number density of 21-cm
absorbers with redshift runs counter to that of the Lyman-\AL\ and
${\rm W}_{\rm r}^{\lambda2796}>1$ \AA\ Mg{\sc \,ii} absorbers
\citep{rtn05}. This is attributed to a significant evolution in either
the CNM filling factor or the covering factor, the former of which
could be the result of a relatively extensive WNM, thus indicating a
higher mean harmonic spin temperature \citep{klm+09}, although
  the larger fractions of warm gas would not exclude the presence of
  cold clumps. Either or both of these scenarios would be tied up in
the $T_{\rm spin}/f$ degeneracy.  Nevertheless, 21-cm detection rates
are significantly lower at high redshift and the filling factor of the
CNM, through either its effect on $T_{\rm spin}$ or $f$, could be a
major contributor to this.

\subsection{21-cm line strength--metallicity correlation}
\label{lsmc}

\begin{figure}
\vspace{6.5cm}
\includegraphics{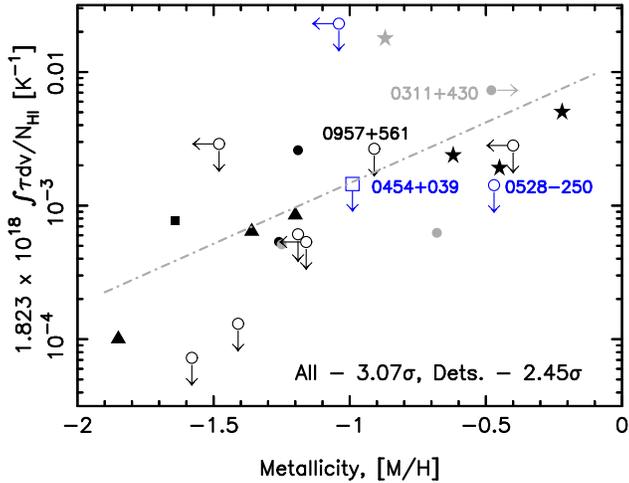}
\caption{The normalised line strength ($\equiv f/T_{\rm spin}$) versus
  the metallicity for the DLAs searched in 21-cm absorption, where the
    symbols are as per Fig.~\ref{Toverf}. The significance of the
  correlation is given for both the whole sample and the detections
  only and the least-squares fit for the detections shown.}
\label{foverT-m}
\end{figure}

Addressing the 21-cm line strength--metallicity correlation ($T_{\rm
  spin}/f$\,--\,[M/H] anti-correlation) of \citet{ctp+07}, the focus of
the 2008 GBT observations (Sect. \ref{GBT08}), it appears that we
should have detected absorption in at least one of the two DLAs for
which reasonable sensitivities were reached (Fig. \ref{foverT-m}).
Regarding the best improved limit\footnote{By a factor of $\approx7$,
  cf. \citealt{cld+96}. See footnote \ref{cld}.}, 21-cm absorption may
not have been detected towards 0528--250, on the grounds that, with an
absorption redshift very close to that of the emission redshift (both
at $z=2.8110$, \citealt{mff04} and references therein), the absorber
is possibly located in the host of the background quasar. We may
therefore expect the neutral gas to be subject to excitation effects
seen in other high redshift 21-cm absorption searches
\citep{cww+08}. However, the presence of H$_2$ in the absorber
suggests that the gas is relatively cool with $T_{\rm kin}\approx 110 - 150$
K (\citealt{spl+05}) and so this is unlikely. 

In any case, referring to Fig.~\ref{foverT-m}, the limit is not overly
low in comparison to 0438--436 (the detection at [M/H]$\,=-0.68$,
$f/T_{\rm spin}=6.3\times10^{-4}$ [K$^{-1}$]) and there is also the
possibility that we have missed the line, with our limit only being
good over the redshift range $z\approx 2.806-2.812$
(Sect. \ref{GBT08}).  Assuming this limit is reliable and including
0311+430, strengthens the correlation
(cf. \citealt{ctp+07}). 

\citet{ykep07} comment that the $z_{\rm abs}=2.289$ absorber towards
0311+430 has an unusually high metallicity, although according to
Fig. \ref{foverT-m}, this has some leeway before the metallicity
becomes atypically high. From its grouping in the line
strength--metallicity plot, we suggest that the absorption is due to a
spiral galaxy, as do \citet{ykep07} on the basis of its low ``spin
temperature''. Support for this assertion is given by its grouping
with the spirals in Fig. \ref{size-z} (the ``unknown'' at
$7\times10^{-3}$ K$^{-1}$, 5 kpc), although this is on the grounds of
a large absorption cross-section, rather than the spin temperature.

\section{Summary}

Through three sets of separate observations, we have undertaken a
survey for 21-cm absorption in a total of twelve DLAs and sub-DLAs.
Due to RFI and low neutral hydrogen column densities, useful
sensitivities were obtained for five of the sample, none of which
resulted in a detection. We add these to the recent results 
from other surveys, which while forming a 
very heterogeneous sample, suggest that, at least statistically:

\begin{enumerate}

\item There appears to be no large difference in the spin
  temperature/covering factor ratio between the low and high redshift
  DLAs (Fig.~\ref{Toverf} and Table \ref{stats}), although the
    non-detections do have a higher mean redshift and more sensitive observations
of these would be required to fully address this. Nevertheless, the fact
    that the mean $T_{\rm spin}/f$ difference is largest for a
    redshift partition of $z_{_{\rm CUT}} = 1.0$ is consistent with
    the geometry effects noted by \citet{cw06}. Other factors, such as
    increasing WNM fractions (increasing $T_{\rm spin}$) or an evolution in
    morphologies (also decreasing $f$) with redshift cannot be ruled
    out, although the $T_{\rm spin}/f$ degeneracy cannot be broken.

  \item The observations since \citet{cmp+03} confirm that the 21-cm line
strength is correlated with the total neutral hydrogen column density,
which further suggests that the spin temperature to covering factor
ratio ($T_{\rm spin}/f$) is {\em roughly} constant, for both
the detections and non-detections (Fig. \ref{tau-N}). 

   \item By combining the radio source sizes with the angular diameter
     distances to the DLAs, we show that the absorber extent
     normalised by the covering factor ($d_{\rm abs}/\sqrt{f}$) is
     correlated with the line-strength for the detections, with the
     non-detections generally exhibiting larger values of $d_{\rm
       abs}/\sqrt{f}$ (Fig. \ref{size-z}). A possible interpretation
     is that the non-detections require larger minimum absorber
     extents in order to be detected. This may also be interpreted as
     the covering factors being lower than for the detections.
             
         This argument relies on the assumption that the unknown
            morphologies of the non-detections are not those of spiral
            galaxies. This is consistent with {\em both} the hypotheses
            of \citet{kc02} and \citet{cmp+03}, in that spiral galaxies
            may have lower spin temperatures or larger covering factors
            than the other morphologies, respectively. The constancy of
            $T_{\rm spin}/f$ between the detections and non-detections
            (point ii, above) may suggest that both contribute, with lower
            covering factors being balanced by lower spin temperatures.
            The absorber extent--line strength correlation further suggests
            that $d_{\rm abs}\propto f^{3/2}/T_{\rm spin}$ for the
            detections. Although, once again, we are limited by the
            spin temperature/covering factor degeneracy, we do believe
            that the large scatter in this ($T_{\rm spin}/f = 60 {\rm
              ~to}\, \gapp 13\,000$~K) is mainly dominated, or at
            least balanced, by different covering factors:
            This represents a vast range of spin temperatures
            (specifically, $T_{\rm spin}\approx200 {\rm ~to}\,\gapp9\,000$~K 
            suggested by \citealt{kc02}). Although having a much smaller
            dynamic range than the spin temperature, the  covering factor
            is subject to three separate effects -- 
           the extent of the absorbing region, the size and morphology of the background
            1.4 GHz emitter, as well as  the effects introduced by the
            geometry of a flat expanding Universe. Furthermore, the beam filling
            of the cold neutral medium in these absorbers could have a significant
            effect on both the covering factor and the spin temperature.
          
    \item We agree with the suspicions of \citet{ykep07} that the $z_{\rm
  abs}=2.289$ absorber towards 0311+430 arises within a spiral galaxy.
  Although not on the basis of its low spin temperature (which we do not
  find unusual at these redshifts), but on its absorption
  cross-section and its grouping with the other spirals on a
  line-strength--metallicity plot.

\end{enumerate}

\section*{Acknowledgments}

We would like to thank the anonymous referee for their prompt and
helpful comments, Ishwara Chandra for his assistance with the
GMRT observations and Jayaram Chengalur for his perseverance in
getting the data to us in a format which {\sc miriad} could
handle. Thanks also to Nadia Lo for her {\sc idl} expertise, which
assisted greatly with the GBT data reduction and to Warren Trotman
(wherever he may be) for the figure of the little telescope (in
Fig.~\ref{dla_schem}), which SJC has used since our Masters write-up
days at Jodrell Bank.

This research has made use of the NASA/IPAC
Extragalactic Database (NED) which is operated by the Jet Propulsion
Laboratory, California Institute of Technology, under contract with
the National Aeronautics and Space Administration. This research has
also made use of NASA's Astrophysics Data System Bibliographic
Service and {\sc asurv} Rev 1.2 \citep{lif92a}, which implements the 
methods presented in \citet{ifn86}.

MTM thanks the
Australian Research Council for a QEII Research Fellowship (DP0877998).


\begin{thebibliography}{}

\bibitem[\protect\citeauthoryear{{Bahcall} \& {Ekers}}{{Bahcall} \&
  {Ekers}}{1969}]{be69}
{Bahcall} J.~N.,  {Ekers} R.~D.,  1969, ApJ, 157, 1055

\bibitem[\protect\citeauthoryear{{Bajtlik}, {Duncan} \& {Ostriker}}{{Bajtlik}
  et~al.}{1988}]{bdo88}
{Bajtlik} S.,  {Duncan} R.~C.,    {Ostriker} J.~P.,  1988, ApJ, 327, 570

\bibitem[\protect\citeauthoryear{{Baker}, {Mathlin}, {Churches} \&
  {Edmunds}}{{Baker} et~al.}{2000}]{bmce00}
{Baker} A.~C.,  {Mathlin} G.~P.,  {Churches} D.~K.,    {Edmunds} M.~G.,  2000,
  in Favata F.,  Kaas A.,   Wilson A.,  eds, Star Formation from the Small to
  the Large Scale, Vol.45 of ESA SP {The Chemical Evolution of the Universe}.
Noordwijk, p.~21

\bibitem[\protect\citeauthoryear{{Beasley}, {Gordon}, {Peck}, {Petrov},
  {MacMillan}, {Fomalont} \& {Ma}}{{Beasley} et~al.}{2002}]{bgp+02}
{Beasley} A.~J.,  {Gordon} D.,  {Peck} A.~B.,  {Petrov} L.,  {MacMillan} D.~S.,
   {Fomalont} E.~B.,    {Ma} C.,  2002, ApJS, 141, 13

\bibitem[\protect\citeauthoryear{{Briggs}, {Brinks} \& {Wolfe}}{{Briggs}
  et~al.}{1997}]{bbw97}
{Briggs} F.~H.,  {Brinks} E.,    {Wolfe} A.~M.,  1997, AJ, 113, 467

\bibitem[\protect\citeauthoryear{Briggs \& Wolfe}{Briggs \& Wolfe}{1983}]{bw83}
Briggs F.~H.,  Wolfe A.~M.,  1983, ApJ, 268, 76

\bibitem[\protect\citeauthoryear{{Carilli}, {Lane}, {de Bruyn}, {Braun} \&
  {Miley}}{{Carilli} et~al.}{1996}]{cld+96}
{Carilli} C.~L.,  {Lane} W.,  {de Bruyn} A.~G.,  {Braun} R.,    {Miley} G.~K.,
  1996, AJ, 111, 1830

\bibitem[\protect\citeauthoryear{Chen \& Lanzetta}{Chen \&
  Lanzetta}{2003}]{cl03}
Chen H.-W.,  Lanzetta K.~M.,  2003, ApJ, 597, 706

\bibitem[\protect\citeauthoryear{Curran}{Curran}{2009}]{cur09a}
Curran S.~J.,  2009, MNRAS, Submitted (arXiv:0910.3998)

\bibitem[\protect\citeauthoryear{Curran, Murphy, Pihlstr\"{o}m, Webb \&
  Purcell}{Curran et~al.}{2005}]{cmp+03}
Curran S.~J.,  Murphy M.~T.,  Pihlstr\"{o}m Y.~M.,  Webb J.~K.,    Purcell
  C.~R.,  2005, MNRAS, 356, 1509

\bibitem[\protect\citeauthoryear{Curran, Tzanavaris, Murphy, Webb \&
  Pihlstr\"{o}m}{Curran et~al.}{2007a}]{ctm+07}
Curran S.~J.,  Tzanavaris P.,  Murphy M.~T.,  Webb J.~K.,    Pihlstr\"{o}m
  Y.~M.,  2007a, MNRAS, 381, L6

\bibitem[\protect\citeauthoryear{Curran, Tzanavaris, Pihlstr\"{o}m \&
  Webb}{Curran et~al.}{2007b}]{ctp+07}
Curran S.~J.,  Tzanavaris P.,  Pihlstr\"{o}m Y.~M.,    Webb J.~K.,  2007b,
  MNRAS, 382, 1331

\bibitem[\protect\citeauthoryear{Curran \& Webb}{Curran \& Webb}{2006}]{cw06}
Curran S.~J.,  Webb J.~K.,  2006, MNRAS, 371, 356

\bibitem[\protect\citeauthoryear{Curran, Webb, Murphy, Bandiera, Corbelli \&
  Flambaum}{Curran et~al.}{2002}]{cwbc01}
Curran S.~J.,  Webb J.~K.,  Murphy M.~T.,  Bandiera R.,  Corbelli E.,
  Flambaum V.~V.,  2002, PASA, 19, 455

\bibitem[\protect\citeauthoryear{Curran, Webb, Murphy \& Carswell}{Curran
  et~al.}{2004}]{cwmc03}
Curran S.~J.,  Webb J.~K.,  Murphy M.~T.,    Carswell R.~F.,  2004, MNRAS, 351,
  L24

\bibitem[\protect\citeauthoryear{Curran, Whiting, Wiklind, Webb, Murphy \&
  Purcell}{Curran et~al.}{2008}]{cww+08}
Curran S.~J.,  Whiting M.~T.,  Wiklind T.,  Webb J.~K.,  Murphy M.~T.,
  Purcell C.~R.,  2008, MNRAS, 391, 765

\bibitem[\protect\citeauthoryear{{Dalgarno} \& {McCray}}{{Dalgarno} \&
  {McCray}}{1972}]{dm72}
{Dalgarno} A.,  {McCray} R.~A.,  1972, Ann. Rev. Astr. Ap., 10, 375

\bibitem[\protect\citeauthoryear{{de Bruyn}, {O'Dea} \& {Baum}}{{de Bruyn}
  et~al.}{1996}]{dob96}
{de Bruyn} A.~G.,  {O'Dea} C.~P.,    {Baum} S.~A.,  1996, A\&A, 305, 450

\bibitem[\protect\citeauthoryear{{Dickey} \& {Lockman}}{{Dickey} \&
  {Lockman}}{1990}]{dl90}
{Dickey} J.~M.,  {Lockman} F.~J.,  1990, ARA\&A, 28, 215

\bibitem[\protect\citeauthoryear{{Douglas}, {Bash}, {Bozyan}, {Torrence} \&
  {Wolfe}}{{Douglas} et~al.}{1996}]{dbb+96}
{Douglas} J.~N.,  {Bash} F.~N.,  {Bozyan} F.~A.,  {Torrence} G.~W.,    {Wolfe}
  C.,  1996, AJ, 111, 1945

\bibitem[\protect\citeauthoryear{Ellison, Yan, Hook, Pettini, Wall \&
  Shaver}{Ellison et~al.}{2001}]{eyh+01}
Ellison S.~L.,  Yan L.,  Hook I.~M.,  Pettini M.,  Wall J.~V.,    Shaver P.,
  2001, A\&A, 379, 393

\bibitem[\protect\citeauthoryear{{Field}}{{Field}}{1959}]{fie59}
{Field} G.~B.,  1959, ApJ, 129, 536

\bibitem[\protect\citeauthoryear{{Fomalont}, {Frey}, {Paragi}, {Gurvits},
  {Scott}, {Taylor}, {Edwards} \& {Hirabayashi}}{{Fomalont}
  et~al.}{2000}]{ffp+00}
{Fomalont} E.~B.,  {Frey} S.,  {Paragi} Z.,  {Gurvits} L.~I.,  {Scott} W.~K.,
  {Taylor} A.~R.,  {Edwards} P.~G.,    {Hirabayashi} H.,  2000, ApJS, 131, 95

\bibitem[\protect\citeauthoryear{{Gupta}, {Srianand}, {Petitjean}, {Noterdaeme}
  \& {Saikia}}{{Gupta} et~al.}{2009}]{gsp+09a}
{Gupta} N.,  {Srianand} R.,  {Petitjean} P.,  {Noterdaeme} P.,    {Saikia}
  D.~J.,  2009, MNRAS, 398, 201

\bibitem[\protect\citeauthoryear{{Gurvits}, {Kellermann} \& {Frey}}{{Gurvits}
  et~al.}{1999}]{gkf99}
{Gurvits} L.~I.,  {Kellermann} K.~I.,    {Frey} S.,  1999, A\&A, 342, 378

\bibitem[\protect\citeauthoryear{{Harvanek}, {Stocke}, {Morse} \&
  {Rhee}}{{Harvanek} et~al.}{1997}]{hsmr97}
{Harvanek} M.,  {Stocke} J.~T.,  {Morse} J.~A.,    {Rhee} G.,  1997, AJ, 114,
  2240

\bibitem[\protect\citeauthoryear{{Isobe}, {Feigelson} \& {Nelson}}{{Isobe}
  et~al.}{1986}]{ifn86}
{Isobe} T.,  {Feigelson} E.,    {Nelson} P.,  1986, ApJ, 306, 490

\bibitem[\protect\citeauthoryear{{Jorgenson}, {Wolfe}, {Prochaska} \&
  {Carswell}}{{Jorgenson} et~al.}{2009}]{jwpc09}
{Jorgenson} R.~A.,  {Wolfe} A.~M.,  {Prochaska} J.~X.,    {Carswell} R.~F.,
  2009, ApJ, 704, 247

\bibitem[\protect\citeauthoryear{Kanekar \& Chengalur}{Kanekar \&
  Chengalur}{2001}]{kc01}
Kanekar N.,  Chengalur J.~N.,  2001, MNRAS, 325, 631

\bibitem[\protect\citeauthoryear{Kanekar \& Chengalur}{Kanekar \&
  Chengalur}{2003}]{kc02}
Kanekar N.,  Chengalur J.~N.,  2003, A\&A, 399, 857

\bibitem[\protect\citeauthoryear{{Kanekar}, {Chengalur} \& {Lane}}{{Kanekar}
  et~al.}{2007}]{kcl06}
{Kanekar} N.,  {Chengalur} J.~N.,    {Lane} W.~M.,  2007, MNRAS, 375, 1528

\bibitem[\protect\citeauthoryear{{Kanekar}, {Lane}, {Momjian}, {Briggs} \&
  {Chengalur}}{{Kanekar} et~al.}{2009a}]{klm+09}
{Kanekar} N.,  {Lane} W.~M.,  {Momjian} E.,  {Briggs} F.~H.,    {Chengalur}
  J.~N.,  2009a, MNRAS, 394, L61

\bibitem[\protect\citeauthoryear{{Kanekar}, {Prochaska}, {Ellison} \&
  {Chengalur}}{{Kanekar} et~al.}{2009b}]{kpec09}
{Kanekar} N.,  {Prochaska} J.~X.,  {Ellison} S.~L.,    {Chengalur} J.~N.,
  2009b, MNRAS, 396, 385

\bibitem[\protect\citeauthoryear{{Kanekar}, {Subrahmanyan}, {Ellison}, {Lane}
  \& {Chengalur}}{{Kanekar} et~al.}{2006}]{kse+06}
{Kanekar} N.,  {Subrahmanyan} R.,  {Ellison} S.~L.,  {Lane} W.,    {Chengalur}
  J.~N.,  2006, MNRAS, 370, L46

\bibitem[\protect\citeauthoryear{{Lanfranchi} \& {Fria{\c c}a}}{{Lanfranchi} \&
  {Fria{\c c}a}}{2003}]{lf03}
{Lanfranchi} G.~A.,  {Fria{\c c}a} A.~C.~S.,  2003, MNRAS, 343, 481

\bibitem[\protect\citeauthoryear{{Lavalley}, {Isobe} \& {Feigelson}}{{Lavalley}
  et~al.}{1992}]{lif92a}
{Lavalley} M.~P.,  {Isobe} T.,    {Feigelson} E.~D.,  1992, in BAAS Vol.~24,
  {ASURV, Pennsylvania State University. Report for the period Jan 1990 - Feb
  1992,} pp 839--840

\bibitem[\protect\citeauthoryear{{Le Brun}, Bergeron, Boiss\'{e} \&
  Deharveng}{{Le Brun} et~al.}{1997}]{lbbd97}
{Le Brun} V.,  Bergeron J.,  Boiss\'{e} P.,    Deharveng J.~M.,  1997, A\&A,
  321, 733

\bibitem[\protect\citeauthoryear{{Lu}, {Sargent}, {Barlow}, {Churchill} \&
  {Vogt}}{{Lu} et~al.}{1996}]{lsb+96}
{Lu} L.,  {Sargent} W. L.~W.,  {Barlow} T.~A.,  {Churchill} C.~W.,    {Vogt}
  S.~S.,  1996, ApJS, 107, 475

\bibitem[\protect\citeauthoryear{{M{\"o}ller}, {Fynbo} \& {Fall}}{{M{\"o}ller}
  et~al.}{2004}]{mff04}
{M{\"o}ller} P.,  {Fynbo} J.~P.~U.,    {Fall} S.~M.,  2004, A\&A, 422, L33

\bibitem[\protect\citeauthoryear{Murphy, Curran \& Webb}{Murphy
  et~al.}{2004}]{mcw04}
Murphy M.~T.,  Curran S.~J.,    Webb J.~K.,  2004, in Duc P.-A.,  Braine J.,
  Brinks E.,  eds, Recycling Intergalactic and Interstellar Matter, IAU
  Symposium No. 217 H$_2$-bearing damped lyman-{$\alpha$} systems as tracers of
  cosmological chemical evolution.
ASP Conf. Ser., San Francisco, p.~252

\bibitem[\protect\citeauthoryear{Murphy, Curran, Webb, M\'{e}nager \&
  Zych}{Murphy et~al.}{2007}]{mcw+07}
Murphy M.~T.,  Curran S.~J.,  Webb J.~K.,  M\'{e}nager H.,    Zych B.~J.,
  2007, MNRAS, 376, 673

\bibitem[\protect\citeauthoryear{{Noterdaeme}, {Ledoux}, {Petitjean} \&
  {Srianand}}{{Noterdaeme} et~al.}{2008}]{nlps08}
{Noterdaeme} P.,  {Ledoux} C.,  {Petitjean} P.,    {Srianand} R.,  2008, A\&A,
  481, 327

\bibitem[\protect\citeauthoryear{Prochaska, Gawiser, Wolfe, Castro \&
  Djorgovski}{Prochaska et~al.}{2003}]{pgw+03}
Prochaska J.~X.,  Gawiser E.,  Wolfe A.~M.,  Castro S.,    Djorgovski S.~G.,
  2003, ApJ, 595, L9

\bibitem[\protect\citeauthoryear{Prochaska \& Herbert-Fort}{Prochaska \&
  Herbert-Fort}{2004}]{ph04}
Prochaska J.~X.,  Herbert-Fort S.,  2004, PASP, 116, 622

\bibitem[\protect\citeauthoryear{Prochaska, Herbert-Fort \& Wolfe}{Prochaska
  et~al.}{2005}]{phw05}
Prochaska J.~X.,  Herbert-Fort S.,    Wolfe A.~M.,  2005, ApJ, 635, 123

\bibitem[\protect\citeauthoryear{Rao, Nestor, Turnshek, Lane, Monier \&
  Bergeron}{Rao et~al.}{2003}]{rnt+03}
Rao S.,  Nestor D.~B.,  Turnshek D.,  Lane W.~M.,  Monier E.~M.,    Bergeron
  J.,  2003, ApJ, 595, 94

\bibitem[\protect\citeauthoryear{Rao, Turnshek \& Nestor}{Rao
  et~al.}{2006}]{rtn05}
Rao S.,  Turnshek D.,    Nestor D.~B.,  2006, ApJ, 636, 610

\bibitem[\protect\citeauthoryear{{Rao}, {Prochaska}, {Howk} \& {Wolfe}}{{Rao}
  et~al.}{2005}]{rphw05}
{Rao} S.~M.,  {Prochaska} J.~X.,  {Howk} J.~C.,    {Wolfe} A.~M.,  2005, AJ,
  129, 9

\bibitem[\protect\citeauthoryear{{Srianand}, {Petitjean}, {Ledoux}, {Ferland}
  \& {Shaw}}{{Srianand} et~al.}{2005}]{spl+05}
{Srianand} R.,  {Petitjean} P.,  {Ledoux} C.,  {Ferland} G.,    {Shaw} G.,
  2005, MNRAS, 362, 549

\bibitem[\protect\citeauthoryear{{Stanghellini}, {Baum}, {O'Dea} \&
  {Morris}}{{Stanghellini} et~al.}{1990}]{sbom90}
{Stanghellini} C.,  {Baum} S.~A.,  {O'Dea} C.~P.,    {Morris} G.~B.,  1990,
  A\&A, 233, 379

\bibitem[\protect\citeauthoryear{{Steidel}, {Bowen}, {Blades} \&
  {Dickenson}}{{Steidel} et~al.}{1995}]{sbbd95}
{Steidel} C.~C.,  {Bowen} D.~V.,  {Blades} J.~C.,    {Dickenson} M.,  1995,
  ApJ, 440, L45

\bibitem[\protect\citeauthoryear{{Tingay}, {Murphy}, {Lovell}, {Costa},
  {McCulloch}, {Edwards}, {Jauncey}, {Reynolds}, {Tzioumis}, {King}, {Jones},
  {Preston}, {Meier}, {van Ommen}, {Nicolson} \& {Quick}}{{Tingay}
  et~al.}{1998}]{tml+98}
{Tingay} S.~J.,  {Murphy} D.~W.,  {Lovell} J.~E.~J., et al.,  1998,
  ApJ, 497, 594

\bibitem[\protect\citeauthoryear{{Turnshek} \& {Bohlin}}{{Turnshek} \&
  {Bohlin}}{1993}]{tb93}
{Turnshek} D.~A.,  {Bohlin} R.~C.,  1993, ApJ, 407, 60

\bibitem[\protect\citeauthoryear{{Turnshek}, {Wolfe}, {Lanzetta}, {Briggs},
  {Cohen}, {Foltz}, {Smith} \& {Wilkes}}{{Turnshek} et~al.}{1989}]{twl+89}
{Turnshek} D.~A.,  {Wolfe} A.~M.,  {Lanzetta} K.~M.,  {Briggs} F.~H.,  {Cohen}
  R.~D.,  {Foltz} C.~B.,  {Smith} H.~E.,    {Wilkes} B.~J.,  1989, ApJ, 344,
  567

\bibitem[\protect\citeauthoryear{{Wolfe} \& {Burbidge}}{{Wolfe} \&
  {Burbidge}}{1975}]{wb75}
{Wolfe} A.~M.,  {Burbidge} G.~R.,  1975, ApJ, 200, 548

\bibitem[\protect\citeauthoryear{{Wolfe}, {Gawiser} \& {Prochaska}}{{Wolfe}
  et~al.}{2005}]{wgp05}
{Wolfe} A.~M.,  {Gawiser} E.,    {Prochaska} J.~X.,  2005, ARA\&A, 43, 861

\bibitem[\protect\citeauthoryear{{Wolfire}, {Hollenbach}, {McKee}, {Tielens} \&
  {Bakes}}{{Wolfire} et~al.}{1995}]{whm+95}
{Wolfire} M.~G.,  {Hollenbach} D.,  {McKee} C.~F.,  {Tielens} A.~G.~G.~M.,
  {Bakes} E.~L.~O.,  1995, ApJ, 443, 152

\bibitem[\protect\citeauthoryear{York, Kanekar, Ellison \& Pettini}{York
  et~al.}{2007}]{ykep07}
York B.~A.,  Kanekar N.,  Ellison S.~L.,    Pettini M.,  2007, MNRAS, 382, L53

\end{thebibliography}

\label{lastpage}
\end{document}